\documentclass[12pt,a4j]{article}
\usepackage[dvips]{graphicx}
\usepackage{amsmath}
\usepackage{amssymb}
\usepackage{enumerate}

\begin{document}
\baselineskip=7mm
\centerline{\bf  Hamiltonian formulation of the extended Green-Naghdi equations}\par
\bigskip
\centerline{Yoshimasa Matsuno\footnote{{\it E-mail address}: matsuno@yamaguchi-u.ac.jp}}\par
\centerline{\it Division of Applied Mathematical Science, Graduate School of Science and Engineering} \par
\centerline{\it Yamaguchi University, Ube, Yamaguchi 755-8611, Japan} \par
\bigskip
\bigskip
\bigskip
\leftline {\bf Abstract} \par
\bigskip
A novel method is developed for extending the Green-Naghdi (GN) shallow-water model  equation to the general system which incorporates the arbitrary higher-order
dispersive effects.
As an illustrative example, we derive a model equation which is accurate to the fourth power of the shallowness parameter while preserving the full nonlinearity of the GN equation, and obtain
its solitary wave solutions by means of a singular perturbation analysis.
We show that  the extended GN equations have the same  Hamiltonian structure 
as that of the GN equation. We also demonstrate that Zakharov's Hamiltonian formulation of surface gravity waves 
is equivalent to that of the extended GN system by rewriting the former system
in terms of the momentum density instead of the velocity potential at the free surface. \par
 \par

\newpage
\leftline{\bf  1. Introduction} \par
\bigskip
 The Green-Naghdi (GN) equation which is also known as the Serre or Su-Gardner equations  models the fully nonlinear and weakly
dispersive surface gravity waves on fluid of finite depth. See Serre [1], Su and Gardner [2] and Green and Naghdi [3]. 
Although the GN equation approximates the Euler equations for the irrotational flows, it exhibits several remarkable
features. In particular,  it has a Hamiltonian formulation which provides a unified framework in exploiting the mathematical structure of various model equations such as
the  Boussinesq, Korteweg-de Vries (KdV) and Camassa-Holm (CH) equations (Camassa and Holm [4] and Camassa {\it et al} [5]). 
A large number of works have been devoted to the studies of  the GN equation from both analytical and numerical points of view. 
A review artile by Barth\'elemy [6]  describes the derivation of the GN equation as well as a method for improving the dispersive effect. 
Furthermore, the improved model equations are tested agianst experiments.
The recent article by Bonneton {\it et al} [7] reviews the high-order numerical methods for
the GN equation and the numerical results in comparison  with breaking random wave propagation experiments.
The following two monographs are concerned with the derivation and
mathematical properties of the GN and other water wave equations: Constantin [8] provides an overwiew of some main results and recent developments in nonlinear water waves including breaking waves and tsunamis.
Lannes [9] addresses the derivation of various asymptotic model equations and their mathematical analysis which is mainly devoted to the well-posedness of the model equations.
   \par
 The GN equation incorporates the  dispersion of  order $\delta^2$, where $\delta=h_0/l$ is the shallowness parameter ($h_0$: mean depth of the fluid, $l$: typical length scale of the wave).
 To improve the dispersion characteristics, various attempts have been made to extend  the GN equation. 
 Among them, the model equations have been derived which include the dispersive terms of  order $\delta^4$ (Kirby [10], Madson and Sch\"affer [11, 12] and Gobbi {\it et al} [13]).
 Numerical computations have been performed for these  equations to examine  the wave profiles and the amplitude-velocity relations as well as 
 the effect of dispersion on the wave characteristics.  Note, however that whether the proposed higher-order dispersive model equations permit the Hamiltonian formulations
 has not been discussed so far. 
 In this paper, we extend the GN equation which is accurate to the dispersive terms of order $\delta^{2n}$ while preserving the full nonlinearlity, where $n$ is an arbitrary
 positive integer. The case $n=1$ corresponds to the GN equation.
 We show that the extended model equations have the same Hamiltonian structure as that of the GN equation. \par
 We consider the two-dimensional 
 irrotational flow of an incompressible and inviscid fluid of uniform depth. 
 The effect of  surface tension is neglected for the sake of simplicity. 
 The governing equation of the water wave problem is given in terms of  the dimensionless variables by
 $$\delta^2\phi_{xx}+\phi_{yy}=0,\quad -\infty<x<\infty,\quad -1<y<\epsilon \eta, \eqno(1.1)$$
 $$\eta_t+\epsilon \phi_x\eta_x={1\over \delta^2}\,\phi_y, \quad y=\epsilon\eta, \eqno(1.2)$$
 $$\phi_t+{\epsilon\over 2\delta^2}(\delta^2\phi_x^2+\phi_y^2)+\eta=0, \quad y=\epsilon \eta, \eqno(1.3)$$
 $$\phi_y=0,\quad y=-1. \eqno(1.4)$$
 Here, $\phi=\phi(x,y,t)$ is the velocity potential, $\eta=\eta(x,t)$ is the profile of the free surface, and the subscripts $x, y,$ and $t$ appended to $\phi$ and
 $\eta$ denote partial differentiations. The dimensional quantities, with tildes, are related to the corresponding dimensionless ones by the relations $\tilde x=lx,
 \tilde y=h_0y, \tilde t=(l/c_0)t, \tilde\eta=a\eta$ and $\tilde\phi=(gla/c_0)\phi$, where $a$ and $c_0$ are characteristic scales of the amplitude and
 velocity of the wave, respectively,  and $g$ is the acceleration due to the gravity. 
  Note that $c_0=\sqrt{gh_0}$ is the long wave phase velocity.
 In the  problem under consideration, one can 
 choose the  two independent dimensionless parameters, $\epsilon=a/h_0$ and $\delta=h_0/l$. The former parameter characterizes the magnitude of  nonlinearity
  whereas the latter characterizes the dispersion or shallowness.  \par
  In Section 2, we provide a recipe for deriving  the model equations. See, for instance Matsuno [14] as for an analogous method which develops a procedure for 
  obtaining the full dispersion model equations of the water wave problem.
   After completing the construction of the extended GN system,
  we derive as an example a model equation which is accurate to order $\delta^4$.
  In Section 3, we show that the extended GN equations can be formulated in  a Hamiltonian form by employing an appropriate Lie-Poisson bracket. 
  At the same time, we demonstrate that the extended equations are equivalent to Zakharov's  equations of motion for surface gravity waves.
  In Section 4, we briefly address the 
  solitary wave solutions of the $\delta^4$ GN model. Finally, Section 5 is devoted to conclusion. \par
  \bigskip
  \leftline {\bf 2. Derivation of the extended Green-Naghdi equations}\par
  \medskip
  \leftline{\it 2.1. The extended GN system}\par
  \medskip
  We first introduce the mean horizontal velocity component $\bar u=\bar u(x,t)$ by
  $$\bar u(x,t)={1\over h}\int_{-1}^{\epsilon\eta}\phi_x(x,y,t)dy, \quad h=1+\epsilon\eta,
   \eqno(2.1)$$
  where $h$ is the total depth of the fluid.   The horizontal and vertical  components of the surface velocity $u$ and $v$ are given respectively by
  $$u(x,t)=\phi_x(x,y,t)|_{y=\epsilon\eta},\eqno(2.2)$$
  $$\quad v(x,t)=\phi_y(x,y,t)|_{y=\epsilon\eta}. \eqno(2.3)$$
  \par
  Multiplying (2.1) by $h$ and differentiating the resultant expression by $x$ and then using (1.1), (1.4), (2.2) and (2.3), we obtain the
  relation $(h\bar u)_x=\epsilon \eta_xu-v/\delta^2$, or since $\epsilon\eta_x=h_x$
  $$v=\delta^2\{-(h\bar u)_x+h_xu\}. \eqno(2.4)$$
  Substitution of (2.4) into (1.2) yields the evolution equation for $h$:
  $$h_t+\epsilon(h\bar u)_x=0. \eqno(2.5) $$
    An advantage of choosing $h$ and $\bar u$ as the dependent variables is that (2.5) becomes
  an exact equation without any approximation.
  \par
  Next, we differentiate (2.2) and (2.3)  by $x$ and $t$ to obtain the relations
  $$u_x=\phi_{xx}+\epsilon\phi_{xy}\eta_x,\quad u_t=\phi_{xt}+\epsilon\phi_{xy}\eta_t, \eqno(2.6)$$
  $$v_x=\phi_{xy}+\epsilon\phi_{yy}\eta_x,\quad v_t=\phi_{yt}+\epsilon\phi_{yy}\eta_t, \eqno(2.7)$$
 where  the derivatives  $\phi_{xx}, \phi_{xy}, \phi_{yy}, \phi_{xt}$ and $\phi_{yt}$ are evaluated at $y=\epsilon\eta$. Similarly, 
 $$\left(\phi_t|_{y=\epsilon\eta}\right)_x=\phi_{xt}+\epsilon\phi_{yt}\eta_x. \eqno(2.8)$$
 Eliminating $\phi_{xt}$ and $\phi_{yt}$ with use of (2.6) and  (2.7), (2.8) becomes
 $$\left(\phi_t|_{y=\epsilon\eta}\right)_x=u_t+v_th_x-v_xh_t. \eqno(2.9)$$
 If we differentiate (1.3) by $x$, insert  (2.4) and (2.9) in the resultant expression and use (2.5), we obtain the evolution equation for $u$:
 $$u_t+v_th_x+\epsilon uu_x+\epsilon h_xuv_x+\eta_x=0. \eqno(2.10)$$
 Using (2.5), Eq. (2.10) can be recast into the form
 $$[h(u+vh_x)]_t+\epsilon[h(u+vh_x)\bar u]_x$$
$$+\epsilon[hv(2h_x\bar u_x+h\bar u_{xx})+h(u-\bar u)(u_x+v_xh_x)]+h\eta_x=0. \eqno(2.11)$$
\par
 The system of equations (2.5) and (2.10) (or (2.11)) is equivalent to the basic Euler system (1.1)-(1.4).
 To obtain the extended GN equations, one needs to express the variables $u$ and $v$  in (2.10) in terms of $h$ and $\bar u$.
 This is always possible as will be exemplified below.
  Consequently, Eq. (2.10) can be written in the form $\bar u_t=\sum_{n=0}^\infty \delta^{2n}K_n$, 
  where $K_n$ are polynomials of $h$ and $\bar u_{nx}, \bar u_{nx,t},\ (\bar u_{nx}=\partial^n\bar u/\partial x^n, n=0, 1, 2 ...).$
 If one retains the terms up to order $\delta^{2n}$, it yields the
 extended GN equation which is accurate to $\delta^{2n}$. 
 In accordance with this fact, we call the system of equations (2.5) and (2.10) (or (2.11)) with   $h$ and $\bar u$ being the dependent variables  the extended GN system. \par
 \bigskip
 \leftline{\it 2.2. The $\delta^4$ model}\par
 \medskip
  Now, we derive the extended GN equation explicitly in the case of $n=2$ by truncating the system of equations (2.5) and (2.11) at order $\delta^4$, which
  we call the $\delta^4$ model.
   To this end, we express the solution of the Laplace equation (1.1) subjected to the boundary condition (1.4) in the form of an infinite series (see, for instance  Whitham [15])
 $$\phi(x,y,t)=\sum_{n=0}^\infty(-1)^n\delta^{2n}{(y+1)^{2n}\over (2n)!}\,f_{2nx}, \quad f_{2nx}={\partial^{2n}f\over \partial x^{2n}}, \eqno(2.12)$$
 where $f=f(x,t)$ is the velocity potential at the fluid bottom $y=-1$.
 The expressions (2.1), (2.2) and (2.3) then become
 $$\bar u=\sum_{n=0}^\infty(-1)^n\delta^{2n}{h^{2n}\over (2n+1)!}\,f_{(2n+1)x}, \eqno(2.13)$$
 $$u=\sum_{n=0}^\infty(-1)^n\delta^{2n}{h^{2n}\over (2n)!}\,f_{(2n+1)x}, \eqno(2.14)$$
  $$v=\sum_{n=1}^\infty(-1)^n\delta^{2n}{h^{2n-1}\over (2n-1)!}\,f_{2nx}. \eqno(2.15)$$
   Retaining the terms of order $\delta^4$,  (2.13)  gives
$$\bar u=f_x-{\delta^2\over 6}h^2f_{xxx}+{\delta^4\over 120}h^4f_{xxxxx}+O(\delta^6). \eqno(2.16)$$
The inverse relation in which $f_x$ is expressed in terms of $h$ and $\bar u$ is achieved by the successive approximation starting from $f_x=\bar u$.
This leads to
$$f_x=\bar u+{\delta^2\over 6}h^2\bar u_{xx}+\delta^4\left\{{h^2\over 36}(h^2\bar u_{xx})_{xx}-{h^4\over 120}\bar u_{xxxx}\right\}
+O(\delta^6). \eqno(2.17)$$
On the other hand, $u$ from (2.14) is expanded as
$$u=f_x-{\delta^2\over 2}h^2f_{xxx}+{\delta^4\over 24}h^4f_{xxxxx}+O(\delta^6). \eqno(2.18)$$
The last step is to introduce $f_x$ from (2.17) into (2.18), giving rise to the relation
$$u=\bar u-{\delta^2\over 3}h^2\bar u_{xx}-\delta^4\left\{{1\over 45}h^4\bar u_{xxxx}+{2\over 9}h^3h_x\bar u_{xxx}+{1\over 18}h^2(h^2)_{xx}\bar u_{xx}\right\}+O(\delta^6). \eqno(2.19)$$
In view of (2.19), the expression of $v$ from (2.4) can be written simply as
$$v=-\delta^2h\bar u_x-{\delta^4\over 3}h^2h_x\bar u_{xx}+O(\delta^6). \eqno(2.20)$$
\par
The evolution equation for $\bar u$ follows by substituting (2.19) and (2.20) into (2.10) and using (2.5) to replace $h_t$.
 After   some manipulations, we arrive at the
compact equation for $\bar u$:
$$\bar u_t+\epsilon\bar u\bar u_x+\eta_x={\delta^2\over 3h}\left\{h^3(\bar u_{xt}+\epsilon\bar u\bar u_{xx}-\epsilon\bar u_x^2)\right\}_x$$
$$+{\delta^4\over 45h}\left[\left\{h^5(\bar u_{xxt}+\epsilon\bar u\bar u_{xxx}-5\epsilon \bar u_x\bar u_{xx})\right\}_x-3\epsilon h^5\bar u_{xx}^2\right]_x + O(\delta^6). \eqno(2.21)$$
If we multiply (2.21) by $h$ and use (2.5), then we can put it in the conservation form
$$(h\bar u)_t+\epsilon\left(h\bar u^2+{h^2\over 2\epsilon^2}\right)_x={\delta^2\over 3}\left\{h^3(\bar u_{xt}+\epsilon\bar u\bar u_{xx}-\epsilon\bar u_x^2)\right\}_x$$
$$+{\delta^4\over 45}\left[\left\{h^5(\bar u_{xxt}+\epsilon\bar u\bar u_{xxx}-5\epsilon \bar u_x\bar u_{xx})\right\}_x-3\epsilon h^5\bar u_{xx}^2\right]_x + O(\delta^6). \eqno(2.22)$$
The system of equations (2.5) and (2.21) (or (2.22)) is the extended version of the GN equation which is accurate to order $\delta^4$ and which retains the full nonlinearity. 
Recall that the amplitude parameter $\epsilon$ which characterizes the magnitude of nonlinearity enters  into $h$ through the relation $h=1+\epsilon\eta$.
What is meant by "full nonlinearity" is that the parameter $\epsilon$ is assumed to be of  order 1. \par
If we use the approximation $h^n=1+\epsilon n\eta +(\epsilon^2)$ in (2.21) under the assumption $\epsilon\ll 1$, where $n$ is a positive integer and
ignore the $\epsilon\delta^2$ and higher-order terms, it reduces to the classical Boussinesq system (Whitham [15])
$$h_t+\epsilon(h\bar u)_x=0,\quad \bar u_t+\epsilon\bar u\bar u_x+\eta_x={\delta^2\over 3}\bar u_{xxt}, \eqno(2.23)$$
 and to  
the original GN system (Serre [1], Su and Gardner [2] and Green and Naghdi [3])
$$h_t+\epsilon(h\bar u)_x=0,\quad \bar u_t+\epsilon\bar u\bar u_x+\eta_x={\delta^2\over 3h}\left\{h^3(\bar u_{xt}+\epsilon\bar u\bar u_{xx}-\epsilon\bar u_x^2)\right\}_x, \eqno(2.24)$$
 if we ignore the $\delta^4$ terms while preserving the full nonlinearity. \par
\medskip
\noindent{\bf Remark 1.} Equation (2.21) involves the nonlinear terms of order $\epsilon^5$. 
If one retains the $\epsilon\delta^2, \epsilon\delta^4$ and $\epsilon^2\delta^2$ terms, then it reduces to a model equation presented in  Madson and Sch\"affer [11] (see Eq. (3.11) therein).
If one imposes  a relation between the parameters $\epsilon$ and $\delta$, then
one can derive the higher-order versions of the various shallow-water model equations. For instance, the scaling $\epsilon=O(\delta^2)\ll 1$ will lead to the higher-order
KdV equation and to the higher-order CH equation under the scaling  $\epsilon=O(\delta)\ll 1$. See Johnson [16] and Constantin and Lannes [17].   \par
\medskip
\noindent{\bf Remark 2.} The linearization of the system of equations (2.5) and (2.21) about the uniform state $h=1$ and $\bar u=0$ yields the system of linear equations for $\eta$ and $\bar u$
$$\eta_t+\bar u_x=0, \quad \bar u_t+\eta_x={\delta^2\over 3}\bar u_{xxt}+{\delta^4\over 45}\bar u_{xxxxt}. \eqno(2.25)$$
Eliminating the variable $\eta$ from this system, one obtains the linear wave equation for $\bar u$
$$\bar u_{tt}-\bar u_{xx}={\delta^2\over 3}\bar u_{xxtt}+{\delta^4\over 45}\bar u_{xxxxtt}. \eqno(2.26)$$
The dispersion relation for equation (2.26) is given by
$$\omega^2={k^2\over 1+{(\delta k)^2\over 3}-{(\delta k)^4\over 45}}. \eqno(2.27)$$
\par
For the wave propagating to the right, the phase velocity $c_p=\omega/k$ and the group velocity $v_g=d\omega/dk$ have the series expansions
$$c_p=1-{1\over 6}(\delta k)^2+{19\over 360}(\delta k)^4-{37\over 2160}(\delta k)^6 +O((\delta k)^8), \eqno(2.28)$$
$$c_g=1-{1\over 2}(\delta k)^2+{19\over 72}(\delta k)^4-{259\over 2160}(\delta k)^6 +O((\delta k)^8). \eqno(2.29)$$
These expressions are compared with the exact dispersion relation $\omega^2=k\,\tanh k\delta/\delta$, as well as the corresponding phase and group velocities which are given respectively by
$$c_p=1-{1\over 6}(\delta k)^2+{19\over 360}(\delta k)^4-{55\over 3024}(\delta k)^6 +O((\delta k)^8), \eqno(2.30)$$
$$c_g=1-{1\over 2}(\delta k)^2+{19\over 72}(\delta k)^4-{55\over 432}(\delta k)^6 +O((\delta k)^8). \eqno(2.31)$$
As expected, both velocities coincide up to  order $(k\delta)^4$. Slight differences appear at order $(k\delta)^6$. \par
\medskip
\noindent{\bf Remark 3.} The system of equations (2.5) and (2.22) exhibits the following three conservation laws:
$$M=\int^\infty_{-\infty}(h-1)dx, \eqno(2.32)$$
$$P=\int^\infty_{-\infty}h\bar u\,dx, \eqno(2.33)$$
$$H={\epsilon^2\over2}\int^\infty_{-\infty}\left[h\bar u^2+ {\delta^2\over 3}h^3\bar u_x^2-{\delta^4\over 45}h^5\bar u_{xx}^2+{1\over\epsilon^2}(h-1)^2\right]dx, \eqno(2.34)$$
which represent the conservation of the mass, momentum and total  energy, respectively. 
The first two conservation laws follow directly from (2.5) and (2.22) whereas the last one can be checked by means of a
straightforward calculation using (2.5) and (2.21). 
A factor $\epsilon^2$ attached in front of the integral sign in (2.34) is only for  convenience.
\par
\bigskip
\leftline{\bf 3. Hamiltonian formulation}\par
\medskip
\leftline{\it 3.1. Hamiltonian structure}\par
\medskip
 Here, we show that the extended GN system (2.5) and (2.11) can be formulated as a Hamiltonian system by introducing an appropriate
Lie-Poisson bracket. 
As pointed out by  Zakharov [18], the basic Euler system of equations (1.1)-(1.4)
 conserves the total energy (i.e., kinetic plus potential energies) 
 $$ H={\epsilon^2\over 2}\int^\infty_{-\infty}\left[\int^{\epsilon\eta}_{-1}\left(\phi_x^2+{1\over\delta^2}\phi_y^2\right)dy+\eta^2\right]dx. \eqno(3.1)$$
 This fact can be confirmed easily using (1.1)-(1.4).
 We substitute (2.12) into (3.1) and then perform the integration with respect to $y$ to express $ H$ in terms of $f_x, f_{xx}, ...$ and $h$. Eliminating $f_x, f_{xx}, ...$ 
 with the aid of the higher-order version of (2.17) yields a series expansion of $H$ in $\delta^2$ as 
 $$ H={\epsilon^2\over 2}\int^\infty_{-\infty}\left[\sum_{n=0}^\infty f_n(h, h_x, h_{xx}, ...; \delta^2)\bar u_{nx}^2+\eta^2\right]dx = \epsilon^2\sum_{n=0}^\infty\delta^{2n} H_n, \eqno(3.2)$$
  where $f_0=h$ and $f_n (\geq 1)$ are polynomials of $h, h_x, h_{xx}, ...\,$($\delta^2$ is simply a parameter).  Specifically, the first four $H_n$ read
 $$ H_0={1\over 2}\int^\infty_{-\infty}\left[h\bar u^2+{1\over\epsilon^2}(h-1)^2\right]dx, \quad
  H_1={1\over 6}\int^\infty_{-\infty}h^3\bar u_x^2dx, \quad
  H_2=-{1\over 90}\int^\infty_{-\infty}h^5\bar u_{xx}^2dx, $$
$$ H_3={1\over 1890}\int^\infty_{-\infty}\left[2h^7\bar u_{xxx}^2-7h^5(hh_x)_x\bar u_{xx}^2\right]dx. \eqno(3.3)$$
\par
Let us now introduce the momentum density $m=m(x,t)$ by
$$\epsilon m={\delta H\over\delta \bar u}, \eqno(3.4)$$
where the operator $\delta/\delta \bar u$ is the variational derivative defined by the relation 
$${\partial\over\partial\epsilon}H[\bar u+\epsilon w]|_{\epsilon=0}=\int^\infty_{-\infty}{\delta H\over\delta \bar u}\,w\, dx, \eqno(3.5)$$
for   arbitrary function $w$.
In terms of $m$ above, the Hamiltonian (or total energy) (3.2) can be put into the simple form
$$H={1\over2}\int^\infty_{-\infty}\left[\epsilon\bar um+(h-1)^2\right]dx. \eqno(3.6)$$
Actually, it follows from (3.2) and (3.4) that
$$m=\epsilon\sum_{n=0}^\infty(-1)^n(f_n\bar u_{nx})_{nx}. \eqno(3.7)$$
Multiplying (3.7) by $\bar u$ and integrating by parts, we obtain
$$\epsilon\int^\infty_{-\infty}\bar um\,dx=\epsilon^2\sum_{n=0}^\infty\int^\infty_{-\infty}f_n(h, h_x, h_{xx}, ...)\bar u_{nx}^2\,dx, \eqno(3.8)$$
which, substituted in (3.2), gives (3.6). \par
To proceed, we regard $m$ and $h$ as the dependent variables instead of $\bar u$ and $h$.
Then, the variation of $H$ in the former variables is found to be
$$\delta H=\int^\infty_{-\infty}\left[\epsilon \bar u\, \delta m
-{\epsilon^2\over 2}\left\{\sum_{n=0}^\infty\sum_{s=0}^{p(n)}(-1)^s\left({\partial f_n\over\partial h_{sx}}\bar u_{nx}^2\right)_{sx}-{2\over\epsilon^2}(h-1)\right\}\delta h\right]dx, \eqno(3.9)$$
where $p(n)$ is a nonnegative positive integer depending on $n$. In particular, $p(0)=p(1)=0$.
To verify (3.9), we modify $H$ from (3.6) as 
$$H=\int^\infty_{-\infty}\left[\epsilon\bar um -{\epsilon\over 2}\bar um +{1\over 2}(h-1)^2\right]dx, \eqno(3.10) $$
and insert the expression of $m$ from (3.7) into the middle term in the integrand, and then take the variations with respect to $m, \bar u$ and $h$. The coefficient of $\delta \bar u$
vanishes by virtue of (3.7), resulting in (3.9).
It readily follows from (3.9) that
$${\delta H\over\delta m}=\epsilon\bar u, \eqno(3.11)$$
$${\delta H\over\delta h}=-{\epsilon^2\over 2}\sum_{n=0}^\infty\sum_{s=0}^{p(n)}(-1)^s\left({\partial f_n\over\partial h_{sx}}\bar u_{nx}^2\right)_{sx}+h-1. \eqno(3.12)$$
\par
\medskip
Now, our main result is given by the following theorem. \par
\medskip
\noindent {\bf Theorem 1.} 
{\it The extended GN system (2.5) and (2.11) can be written in the following Hamiltonian form
$$\begin{pmatrix} m_t \\  h_t\end{pmatrix}=-\begin{pmatrix} \partial_xm+m\partial_x  & h\partial_x  \\ \partial_x h & 0\end{pmatrix}
 \begin{pmatrix} {\delta H/\delta m}\\ {\delta H/\delta h} \end{pmatrix}, \eqno(3.13)$$
where the derivative $\partial_x\equiv \partial/\partial x$  operates on all terms it multiplies to the right.
Introduce the Lie-Poisson bracket between any pair of smooth  functionals $F$ and $G$
$$\{F,G\}=-\int^\infty_{-\infty}\left[{\delta F\over \delta m}(m\partial_x+\partial_x m){\delta G\over \delta m}+{\delta F\over \delta m}\,h\partial_x{\delta G\over \delta h}
+{\delta F\over \delta h}\,\partial_x h{\delta G\over \delta m}\right]dx. \eqno(3.14)$$
Then, the equations of motion for $m$ and $h$ from (3.13)
 are expressed as
 $$m_t=\{m,H\},\quad  h_t=\{h,H\}. \eqno(3.15)$$  }\par
 \medskip
{\noindent {\bf Proof.} The proof of  Theorem 1  relies on the following two relations:
$$m=\epsilon h(u+vh_x), \eqno(3.16)$$
$$m\bar u_x+{h\over\epsilon}\left({\delta H\over\delta h}\right)_x=\epsilon[hv(2h_x\bar u_x+h\bar u_{xx})+h(u-\bar u)(u_x+v_xh_x)]+h\eta_x, \eqno(3.17)$$
which we shall address first and then proceed to the proof of  Theorem 1. \par
 We modify the Hamiltonian $H$ from (3.1). First, use (1.1) to obtain
$$\phi_x^2+{\phi_y^2\over\delta^2}=(\phi_x\phi)_x+{1\over\delta^2}(\phi_y\phi)_y. $$
If we introduce this expression into (3.1) and perform the integration with respect to $y$ under  the boundary condition (1.4) as well as  (2.2) and (2.3), we deduce
$$H={\epsilon^2\over 2}\int^\infty_{-\infty}\left[\left(-\epsilon \eta_xu+{v\over\delta^2}\right)\psi+\eta^2\right]dx, $$
where $\psi$ is the velocity potential at the free surface given by
$$\psi=\phi(x,\epsilon\eta,t). \eqno(3.18)$$
By virtue of (2.4), the above Hamiltonian reduces, after integrating by parts, to
$$H={\epsilon^2\over 2}\int^\infty_{-\infty}\left[h\bar u\psi_x+\eta^2\right]dx. $$
Equating this expression with (3.6), we arrive at the integral identity
$$\int^\infty_{-\infty}\bar u(m-\epsilon h\psi_x)dx=0.$$
Since this must hold for arbitrary $\bar u$, we obtain an important relation
$$m=\epsilon h\psi_x, \eqno(3.19)$$
which connects the variable $\psi_x$ with the variable $m$.
It follows by differentiating (3.18) by $x$ and using (2.2) and (2.3) that
$$\psi_x=\phi_x(x,\epsilon\eta,t)+\epsilon\phi_y(x,\epsilon\eta,t)\eta_x=u+vh_x,$$
which, coupled with (3.19), yields (3.16).\par
 We substitute $m$ from (3.16) into (3.17), divide the resultant expression by $h$ and use (2.4). The relation thus obtained can be integrated by $x$ to give
$${1\over\epsilon^2}{\delta H\over\delta h}={1\over 2}u^2+{v^2\over 2\delta^2}+hv\bar u_x-u\bar u+{1\over\epsilon^2}(h-1). \eqno(3.20)$$
By taking the variational derivative of $H$ from (3.1) with respect to $h$, we obtain
$${\delta H\over\delta h}={\epsilon^2\over 2}\left(u^2+{v^2\over\delta^2}+{2\eta\over\epsilon}\right)
+\epsilon^2\int^\infty_{-\infty}\left(-\epsilon\eta_xu+{v\over\delta^2}\right){\delta\phi\over\delta h}\Big |_{y=\epsilon\eta}dx. $$
The following relation comes from the formula $\delta h(x,t)/\delta h(x^\prime,t)=\delta(x-x^\prime)$, where $\delta(x-x^\prime)$ is Dirac's delta function:
$${\delta\phi(x,y,t)\over\delta h(x^\prime,t)}\Bigg |_{y=\epsilon\eta}={\delta\psi(x,t)\over\delta h(x^\prime,t)}-v\delta(x-x^\prime).$$
We use  this relation and (2.4) in the expression of $\delta H/\delta h$ to obtain
$${\delta H\over\delta h}={\epsilon^2\over 2}\left(u^2+{v^2\over\delta^2}+{2\eta\over\epsilon}\right)+\epsilon^2(h\bar u)_xv+\epsilon^2\int^\infty_{-\infty}h\bar u{\delta \psi_x\over\delta h}dx.$$
Last, substituting the above expression into the left-hand side of (3.20), the relation to be proved reduces to
$$(h\bar u)_xv + \int^\infty_{-\infty}h\bar u{\delta \psi_x\over\delta h}dx=hv\bar u_x-u\bar u. \eqno(3.21)$$
The integral on the left-hand side of (3.21)  can be evaluated by using the relation $\psi_x=m/\epsilon h$ from (3.19), which leads to
$$\int^\infty_{-\infty}h\bar u{\delta \psi_x\over\delta h}dx=-{m\bar u\over \epsilon h}=-(u+vh_x)\bar u,$$
showing that (3.21) holds identically.  Hence, we establish (3.17). \par
Now,   Theorem 1 follows immediately from (3.16) and (3.17). First, we note that
 the introduction of (3.11) into (3.13) yields the system of evolution equations for $m$ and $h$:
$$m_t=-\epsilon(\bar u m)_x-\epsilon m\bar u_x-h\left({\delta H\over\delta h}\right)_x, \eqno(3.22)$$
$$h_t=-\epsilon(h\bar u)_x. \eqno(3.23)$$
Equation (3.23) is just Eq. (2.5). If we substitute (3.16) and (3.17) into (3.22), we see that the resultant equation is in agreement  with  Eq. (2.11).
It is obvious that (3.15) with (3.14) is equivalent to (3.13).
 \quad $\square$ \par
\medskip
\noindent{\bf Remark 4.} The bracket (3.14) has been used in Holm [19] to verify the Hamiltonian structure of two-dimensional shallow-water  hydrodynamics with nonlinear dispersion, 
 as well as that of the GN equation which has been detailed by Constantin [20]. It has a skew-symmetry $\{F,G\}=-\{G,F\}$  
and satisfies the Jacobi identity $\{F,\{G,H\}\}+\{G,\{H,F\}\}+\{H,\{F,G\}\}=0$ (Constantin [20]). \par 
 \medskip
 \noindent{\bf Remark 5.} If we retain the terms of order $\delta^4$, then $m$ and $\delta H/\delta h$  are found to be
 $$m=\epsilon\left[h\bar u-{\delta^2\over 3}(h^3\bar u_x)_x-{\delta^4\over 45}(h^5\bar u_{xx})_{xx}\right], \eqno(3.24)$$
 $${\delta H\over\delta h}=-{\epsilon^2\over 2}\left\{\bar u^2+\delta^2h^2\bar u_x^2-{\delta^4\over 9}h^4\bar u_{xx}^2-{2\over\epsilon^2}(h-1)\right\}. \eqno(3.25)$$
Equation (3.22) then becomes
$$m_t=-\epsilon\bar um_x-2\epsilon\bar u_xm+{\epsilon^2\over 2}h\left[\bar u^2+\delta^2h^2\bar u_x^2-{\delta^4\over 9}h^4\bar u_{xx}^2-{2\over\epsilon^2}(h-1)\right]_x. \eqno(3.26)$$
One can check by a direct computation that (3.26) indeed reproduces (2.22).
\par
\medskip
 \noindent{\bf Remark 6.} In addition to  the conservation of mass, momentum and total energy which are given respectively by (2.32), (2.33) and (2.34), the extended GN
 system exhibits the fourth conservation law $L\equiv \int^\infty_{-\infty}m/h\,dx$. This immediately follows from the equation $(m/h)_t=-(\epsilon \bar um/h+\delta H/\delta h)_x$
 which is derived from the system of equations (3.22) and (3.23). In view of (3.16), the density of $L$ is written as $m/h=\epsilon(u+vh_x)$. Since the
 unit tangent vector ${\bf t}$ to the free surface is given by ${\bf t}=(1/\sqrt{1+h_x^2}, h_x/\sqrt{1+h_x^2})$ and the surface velocity by ${\bf u}=(u, v)$, the density $m/h$ is equal to the
 tangential component $u_t\equiv {\bf u}\cdot {\bf t}$ of the surface velocity multiplied by a factor $\epsilon \sqrt{1+h_x^2}$.  It  turns out that $L$ is represented by the line integral of the tangent velocity
 along the free surface. To be more specific, $L=\epsilon\int_Cu_tds$, where
 $ds=\sqrt{1+h_x^2}\,dx$ is the line element of the curve $C$ representing the free surface. 
 Using (3.19), this integral can be evaluated explicitly as $L=\epsilon(\psi_+-\psi_-)$ with 
 $\psi_\pm\equiv {\rm lim}_{x\rightarrow \pm\infty}\psi(x, t)$, 
 showing that the conserved quantity $L$ is determined only by the limiting values $\psi_\pm$  of the velocity potential at the free surface.
 Note also that the quantity $\int^\infty_{-\infty}mdx$ is conserved. However, it is not independent of the momentum $P$ given by (2.33).
 Actually, integrating the relation (3.7) with $f_0=h$, the above conserved quantity reduces to  $\epsilon\int^\infty_{-\infty}h\bar u\,dx=\epsilon P$. \par
 \medskip
\leftline{\it 3.2. Relation to Zakharov's Hamiltonian formulation}\par
\medskip
  The water wave problem posed by (1.1)-(1.4) permits a Hamiltonian formulation.  Following Zakharov [18],  the equations of motion   for surface gravity waves  can be written 
in terms of the variables $h$ and $\psi_x$ as 
$$h_t=-{1\over\epsilon}\left({\delta H\over\delta\psi_x}\right)_x, \eqno(3.27)$$
$$\psi_{xt}=-{1\over\epsilon}\left({\delta H\over\delta h}\right)_x. \eqno(3.28)$$
Note that the variable $\psi_x$ is used instead of $\psi$. This choice  is suitable for the present analysis.
If we define the bracket between any pair of smooth  functionals $F$ and $G$ by
$$\{F,G\}=-{1\over \epsilon}\int^\infty_{-\infty}\left[{\delta F\over \delta h}\left({\delta G\over \delta \psi_x}\right)_x-\left({\delta F\over \delta \psi_x}\right)_x{\delta G\over \delta h}
\right]dx, \eqno(3.29)$$
then, Eqs. (3.27) and (3.28) are expressed as
$$h_t=\{h, H\}, \quad \psi_{xt}=\{\psi_x, H\}. \eqno(3.30)$$
\par
Under the above setting, the following theorem is established: \par
\bigskip
\noindent {\bf Theorem 2.}\ {\it The system of equations (3.13) is  equivalent to Zakharov's system of equations (3.27) and (3.28).} \par
\medskip
\noindent{\bf Proof.} First, we change the variables $h$ and $\psi_x$ to $h$ and $m$ and rewrite Eqs. (3.27) and (3.28) in terms of the latter variables. 
To this end, we evaluate the variation  of the Hamiltonian $H$ in two alternative ways:
$$\int^\infty_{-\infty}\left[{\delta H\over\delta h}\bigg |_{\psi_x:{\rm fixed}}\delta h+{\delta H\over\delta \psi_x}\bigg |_{h:{\rm fixed}}\delta \psi_x\right]dx
=\int^\infty_{-\infty}\left[{\delta H\over\delta h}\bigg |_{m:{\rm fixed}}\delta h+{\delta H\over\delta m}\bigg |_{h:{\rm fixed}}\delta m\right]dx. \eqno(3.31)$$
We compute the variation of $m$ from (3.19) as $\delta m=\epsilon\psi_x\delta h+\epsilon h\delta\psi_x$ and subsitute this expression into the right-hand side of (3.31) and then
compare the coefficient of $\delta h$ and $\delta\psi_x$ on both sides. It follows that
$${\delta H\over\delta h}\bigg |_{\psi_x:{\rm fixed}}={\delta H\over\delta h}\bigg |_{m:{\rm fixed}}+\epsilon\,{\delta H\over\delta m}\bigg |_{h:{\rm fixed}}\psi_x, \eqno(3.32)$$
$${\delta H\over\delta \psi_x}\bigg |_{h:{\rm fixed}}=\epsilon h\,{\delta H\over\delta m}\bigg |_{h:{\rm fixed}}. \eqno(3.33)$$
Taking into account  (3.11) and (3.19), the above relations become
$${\delta H\over\delta h}\bigg |_{\psi_x:{\rm fixed}}={\delta H\over\delta h}\bigg |_{m:{\rm fixed}}+{\epsilon\bar u m\over h}, \eqno(3.34)$$
$${\delta H\over\delta \psi_x}\bigg |_{h:{\rm fixed}}=\epsilon^2 h\bar u.  \eqno(3.35)$$
\par
These formulas enable us to transform the Zakharov system to the extended GN system.
Actually, Eq. (3.27) with (3.35)  yields Eq. (3.23).  On the other hand, introducing (3.19) and (3.34) into  (3.28), we obtain
$$\left({m\over h}\right)_t={m_t\over h}-{mh_t\over h^2}=-\left({\delta H\over\delta h}\right)_x-\epsilon\left({\bar u m\over h}\right)_x. \eqno(3.36)$$
where we have used the notation $(\delta H/\delta h)|_{m:{\rm fixed}}=\delta H/\delta h$ for simplicity.
If we replace $h_t$ in the middle expression of (3.36) by the right-hand side of Eq. (3.23), then the  equation reduces to  Eq. (3.22) after multiplying $h$ on both sides.
Last, we use (3.32) and (3.33) with $H$ being replaced by $F$ and $G$, respectively and (3.19) for $\psi_x$ to rewrite the bracket (3.29) in terms of $h$ and $m$. The resultant
expression becomes
$$\{F, G\}=-\int_{-\infty}^\infty\left[\left({\delta F\over \delta h}+{m\over h}{\delta F\over \delta m} \right)\left(h\,{\delta G\over \delta m}\right)_x
-\left(h\,{\delta F\over \delta m}\right)_x\left({\delta G\over \delta h}+{m\over h}{\delta G\over \delta m} \right)\right]dx. \eqno(3.37)$$
Integrating the second term of (3.37) by parts and rearranging the integrand, we confirm that (3.37) transforms to the bracket defined by (3.14).
 \quad $\square$ \par
\par 
\bigskip
\leftline {\bf 4. Solitary wave solutions}\par
\medskip
We seek the solitary wave solutions of the extended GN system (2.5) and (2.22) in the form of the traveling wave $h=h(\xi), \bar u=\bar u(\xi), (\xi=x-ct)$.
We impose the boundary conditions $h\rightarrow 1, h^\prime \rightarrow 0 $ 
and $ \bar u, \bar u^{\prime},
\bar u^{\prime\prime}, \bar u^{\prime\prime\prime}, \bar u^{\prime\prime\prime\prime}\rightarrow 0$ as $|x|\rightarrow \infty$, where $c(>0)$ is the
velocity of the solitary wave and  primes refer to  differentiation with respect to $\xi$. 
If we introduce these forms into (2.5) and (2.22) and integrate each equation once with respect to $\xi$ while taking into account the boundary conditions, we deduce 
$$-c(h-1)+\epsilon h\bar u=0, \eqno(4.1)$$
$$-ch\bar u+\epsilon\left(h\bar u^2+{h^2-1\over 2\epsilon^2}\right)
={\delta^2\over 3}h^3(-c\bar u^{\prime\prime}+\epsilon \bar u\bar u^{\prime\prime}-\epsilon (\bar u^\prime)^2)$$
$$+{\delta^4\over 45}\left[\left\{h^5(-c\bar u^{\prime\prime\prime}+\epsilon \bar u\bar u^{\prime\prime\prime}-5\epsilon\bar u^{\prime}\bar u^{\prime\prime})\right\}^\prime
-3\epsilon h^5(\bar u^{\prime\prime})^2\right]. \eqno(4.2)$$
We use (4.1) to express $\bar u$ in terms of $h$ as
$$\bar u={c\over\epsilon}\left(1-{1\over h}\right). \eqno(4.3)$$
Substituting (4.3) into (4.2), multiplying the resultant equation by $h^\prime/h^2$ and then integrating, we obtain the third-order  nonlinear differential  equation for $h$:
$$h^3-(c^2+2)h^2+(2c^2+1)h-c^2$$
$$=-{\delta^2c^2\over 3}(h^\prime)^2-{\delta^4c^2\over 45}\left[2h^2h^\prime h^{\prime\prime\prime}-h^2(h^{\prime\prime})^2
+2h(h^\prime)^2h^{\prime\prime}-12(h^\prime)^4\right]. \eqno(4.4)$$
\par
If we neglect the $\delta^4$ terms on the right-hand side of (4.4), then it reduces to the corresponding equation for the GN equation. 
Unlike the Boussinesq system (2.23), it exhibits the analytical solitary wave solution
of the following form which has been obtained for the first time by Serre [1] and later by Su and Gardner [2]:
$$h=1+(c^2-1)\,{\rm sech}^2{\sqrt{3(c^2-1)}\over 2c\delta}\xi, \quad (c>1). \eqno(4.5)$$
An inspection reveals, however  that Eq. (4.4) would not have analytical  solutions and hence numerical analysis may be necessary to extract the characteristics associated with
the solitary waves such as the surface profile and the velocity-amplitude relation.  Instead, we perform a singular perturbation analysis. \par
Note first that the parameter $\delta$ in Eq. (4.4) disappears  by means of the scaling $\xi\rightarrow \delta\xi$. Furthermore, the form of the solitary wave solution (4.5) suggests that
an appropriate scaling of the variable $\xi$ would be $b\zeta$, where $\zeta=\xi/\delta$ and the unknown parameter $b$ plays the role of the wavenumber which is determined in the course of 
the pertubation analysis. In accordance with this observation,
 we expand $h$ and $c^2$ in powers of $\epsilon$ as
$$h(b\zeta)=1+\epsilon h_1+\epsilon^2h_2+\epsilon^3h_3+O(\epsilon^4), \eqno(4.6)$$
$$c^2=1+\epsilon c_1+\epsilon^2c_2+\epsilon^3c_3+O(\epsilon^4), \eqno(4.7)$$
with
$$b^2=\epsilon b_1+\epsilon^2b_2+\epsilon^3b_3+O(\epsilon^4). \eqno(4.8)$$
 We employ the  conditions $h_1(0)=1$ and $h_j(0)=0, j\ge 2$ which define the amplitude $\epsilon$ by $h(0)=1+\epsilon$.
The method for constructing solitary wave solutions is well-known. See, for instance Fenton [21]. Hence, we summarize the result. \par
The solution to the boundary value problem includes the secular term like $\zeta\,{\rm sech}^2b\zeta\tanh b\zeta$. 
To ensure the uniform validity of the expansion (4.6), one must eliminate the secular terms by
choosing the coefficients $b_j$ and $c_j\ (j=1, 2, ...)$ appropriately. The solitary wave solution thus constructed can be written as
$$h(b\zeta)=1+\epsilon\, {\rm sech}^2b\zeta-{3\over 4}\epsilon^2{\rm sech}^2b\zeta\tanh^2b\zeta
+{5\over 8}\epsilon^3\left(1-{149\over 50}{\rm sech}^2b\zeta\right){\rm sech}^2b\zeta\tanh^2b\zeta+O(\epsilon^4), \eqno(4.9)$$
$$c^2=1+\epsilon-{1\over 20}\epsilon^2-{1\over 10}\epsilon^3+O(\epsilon^4), \eqno(4.10)$$
$$b=\sqrt{{3\over 4}\epsilon}\left(1-{5\over 8}\epsilon+{71\over 128}\epsilon^2+O(\epsilon^3)\right). \eqno(4.11)$$
The above solution coincides with the third-order solution obtained in Fenton [21] and Grimshaw [22] up to order $\epsilon^2$. The discrepancy appears at order $\epsilon^3$.
The reason is obvious since the latter takes into account the $\delta^6$ terms 
which have been neglected in our model equation. To be more specific, the scaling $\epsilon=O(\delta^2)$ is assumed in Fenton [21] and Grimshaw [22] 
so that the  $\delta^6$ terms turn out to be comparable with the $\epsilon^3$ terms.\par
\medskip
\noindent {\bf Remark 7.} Introduce the  quantity $Q$ by
$$Q= \int^\infty_{-\infty}m\left(1-{1\over h}\right)dx, \eqno(4.12)$$
which has been shown to be  conserved by the extended GN system (See Remark 6).
For the $\delta^4$ model, the solitary wave solution $h(\xi)$ and $m(\xi)$ is a critical point of the functional $H-cQ$ with respect to $m$ and $h$. Specifically,
$${\delta H\over\delta m}-c\,{\delta Q\over\delta m}=0, \quad {\delta H\over\delta h}-c\,{\delta Q\over\delta h}=0. \eqno(4.13)$$
This fact can be confirmed by a direct computation using (3.11), (3.24), (3.25) and (4.4). 
Thus, a variational characterization is possible for the solitary wave solution of the $\delta^4$ model, even if it does not exhaibit exact solitary
wave solutions. It will be an important issue to establish this assertion for the general $\delta^{2n}\,(n\geq 3)$ model.
In the case of the GN equation, the similar result has been pointed out
and used to prove the linear stability of the solitary wave solution (4.5) (Li [23]).
 \par
\bigskip
\leftline{\bf 5. Conclusion} \par
\medskip
 We have presented a novel method  for extending the GN equation to the general system which incorporates the arbitrary higher-order dispersive terms while preserving the full nonlinearity.
 As an illustrative example,  we have derived a model equation which is accurate to  order $\delta^4$ for which
 the perturbation analysis  reproduces the known solitary wave solutions at the level of  order $\epsilon^2$.
 Nevertheless, the numerical analysis will be necessary to extract the fully nonlinear nature of the model equation, as has been performed for the GN equation by Li {\it et al} [24] and 
 Mitsotakis {\it et al} [25]. 
 We have shown that our extended system permits the same Hamiltonian structure as that of the GN equation. 
 In the process of establishing this fact,  the energy integral of the basic Euler system has been used efficiently. We have also verified that Zakharov's Hamiltonian formulation of the
 current water wave problem is equivalent to that of the extended GN equations. 
 This has been accomplished by rewriting Zakharov's equations 
 in terms of the total depth of the fluid and momentum density. 
 The extension of our model equations to the more general setting with an uneven bottom topography and its three-dimensional generalization
 will be done along with the same  procedure as that developed here.
In any case, there are many things to be resolved by future study. \par
\bigskip
\newpage
\leftline{\bf Reference} \par
\baselineskip=5.5mm
\begin{enumerate}[{[1]}]
\item  F. Serre,  Contribution \`a l'\'{e}tude des \'{e}coulements permanents et variables dans les canaux,   Houille Blanche 3(1953) 374-388.
\item  C.H. Su, C.S. Gardner,   Korteweg-de Vries equation and generalizations. III. Derivation of the Korteweg-de Vries equation and Burgers equation,
  J. Math. Phys.   10(1969) 536-539. 
\item  A.E. Green, P.M.  Naghdi,    A derivation of equations for wave propagation in water of variable depth,  J. Fluid Mech. 78(1976)  237-246. 
\item  R. Camassa, D.D. Holm,    An integrable shallow water equation with peaked solitons,  Phys. Rev. Lett.  71(1993) 1661-1664. 
\item  R. Camassa, D.D.  Holm,  J.M. Hyman,  A new integrable shallow water equation,   Adv. Appl. Mech. 31(1994) 1-33.
\item  E. Barth\'elemy,  Nonlinear shallow water theories for coastal waves,  Surv.  Geophys. 25(2004) 315-337. 
\item  P. Bonneton,   E. Barthelemy,  F.  Chazel,  R. Cienfuegos,  D. Lannes, F. Marche, M. Tissier,   
Recent advances in Serre-Green Naghdi modelling for wave transformation,
 breaking and runup processes,   Euro. J. Mech. B/Fluids  30(2011) 589-597. 
 \item A. Constantin,  Nonlinear Water Waves with Applications to Wave-Current Interactions and Tsunamis, SIAM, Philadelphia, 2011.
  \item D.  Lannes,    Water waves problem: Mathematical analysis and asymptotics,  American Mathematical Society, 2013.
\item  J.T. Kirby, Nonlinear dispersive long waves in water of variable depth,
in: J.N. Hunt(Ed.),
  Advances in Fluid Mechanics 10, Computational Mechanics Publ., Boston, 1999. pp.  55-125.
\item  P.A. Madsen, H.A. Sch\"affer,   Higher-order Boussinesq-type equations for surface gravity waves: derivation and analysis,
 Phil. Trans. R. Soc. Lond. A   356(1998)  3123-3184.
\item  P.A. Madsen, H.A. Sch\"affer, A review of Boussinesq-type equations for gravity waves,
  in: P. Liu(Ed.) Advances in Coastal and Ocean Engineering 5,  World Scientific, Singapore, 1999, pp. 1-94. 
\item  M.F. Gobbi, J.T. Kirby, G. Wei,   A fully nonlinear Boussinesq model for surface waves. Part 2. Extention to $O(kh)^4$,
  J. Fluid Mech.   405(2000) 181-210. 
\item Y.  Matsuno,  Nonlinear evolutions of surface gravity waves on fluids of finite depth, Phys. Rev. Lett.  69(1992) 609-611. 
\item  G.B. Whitham,  Linear and Nonlinear Waves. Wiley, New York, 1974. 
\item R.S. Johnson,  Camassa-Holm, Korteweg-de Vries and related models for water waves, J. Fluid Mech. 455(2002) 63-82.
\item  A. Constantin, D.  Lannes,  The hydrodynamic relevance of the Camassa-Holm and Degasperis-Procesi equations,   Arch. Rational Mech. Anal.  192(2009) 165-186.
\item  V.E. Zakharov, Stability of periodic waves of finite amplitude on the surface of a deep fluid, J. Appl. Mech. Tech. Phys.  9(1968) 190-194. 
\item  D.D. Holm,  Hamiltonian structure for two-dimensional hydrodynamics with nonlinear dispersion, Phys. Fluids   31(1988) 2371-2373. 
\item  A. Constantin,    The Hamiltonian structure of the Camassa-Holm equation, Exp. Math. 15(1997) 53-85. 
\item  J. Fenton,    A ninth-order solution for the solitary wave, J. Fluid Mech. 53(1972) 257-271. 
\item  R. Grimshaw,  The solitary wave in water of variable depth. Part 2,  J. Fluid Mech. 46(1971) 611-622. 
\item Y.A. Li,  Linear stability of solitary waves of the Green-Naghdi equations,  Comm. Pure Appl. Math. 54(2001) 501-536. 
\item  Y.A. Li, J.M.  Hyman,  W.   Choi,    A numerical study of of the exact evolution equations for surface waves in water of finite depth,  Stud. Appl. Math. 113(2004) 303-324. 
\item  D. Mitsotakis, B. Ilan, D. Dutykh,   On the Galerkin/finite-element method for the Serre equations,   J. Sci. Comp. 61(2014) 166-195. 
\end{enumerate}

\end{document}